\documentclass[12pt]{article}

\usepackage{amsfonts, amsmath, amssymb, bm, bbm, bbold, latexsym}
\usepackage{esint} 

\usepackage{graphicx}
\usepackage{xcolor}
\usepackage{hyperref}
\usepackage{slashed} 

\usepackage{import}

\usepackage{enumerate}

\usepackage{physics}

\usepackage[lmargin=1.0in, rmargin=1.0in, tmargin=1.0in, bmargin=1.0in]{geometry}

\usepackage{cancel}

\usepackage{empheq}

\usepackage{comment}

\usepackage{setspace}
\begin{document}



\title{\bf Geometric Structure of Defects in General Teleparallel Geometry}
\author{  Muzaffer Adak$^1$, Ertan Kok$^1$, Mehmet Orhan$^2$  \\
  {\small $^1$Computational and Gravitational Physics Laboratory,}\\ {\small Department of Physics, Faculty of Science, Pamukkale University, Denizli, Türkiye} \\
  {\small $^2$Department of Mechanical Engineering, Faculty of Engineering, } \\
  {\small Pamukkale University, Denizli, Türkiye} \\
      {\small {\it E-mail:} {madak@pau.edu.tr, ekok@pau.edu.tr, morhan@pau.edu.tr}}}

  \vskip 1cm
\date{November 24, 2026}
\maketitle
\thispagestyle{empty}
\begin{abstract}
 \noindent
We develop a curvature-free geometric framework for the description of topological defect structures. 
In conventional metric-affine models, dislocations are associated with torsion, disclinations with curvature, and point defects with non-metricity. However, this identification leads to a structural overlap in the defect classification, while curvature-squared metric formulations may also involve higher-order field equations. 

In this work we reformulate the defect variables within general teleparallel geometry, characterized by the vanishing of the total curvature, $R^a{}_b = 0$, while allowing nontrivial torsion and non-metricity. Within this setting, a reduced set of effective defect variables is constructed from selected components of the non-Riemannian field strengths. The torsion trace is adopted as an effective Burgers-type variable, the second trace of non-metricity as an effective Frank-type variable, and the first trace of non-metricity as a point-defect variable, while the scalar torsion component remains as an additional independent sector.

The resulting structure defines a constrained geometric field system that admits a metric formulation. Quadratic Lagrangians constructed from torsion and non-metricity lead to second-order field equations in the teleparallel metric
formulation considered here, without requiring curvature-squared terms. The framework provides a reduced geometric classification of effective defect fields and clarifies their underlying variational structure.
  \\



\end{abstract}

\section{Introduction}

Non-Riemannian geometry provides a natural arena for the formulation of field theories in which the metric and the affine connection are treated as independent variables. In such geometries, torsion and non-metricity arise as field strengths associated with the affine structure, and their algebraic and differential identities encode structural constraints of the underlying system. When curvature is required to vanish while torsion and non-metricity remain nontrivial, $R^a{}_b = 0$, one obtains a general teleparallel field system whose dynamical content is entirely carried by non-Riemannian degrees of freedom.

Geometric descriptions of material defects offer a concrete realization of non-Riemannian structures. Since the early works of Kondo, Nye, Bilby, and Kröner \cite{kondo-1952}-\cite{kroner-1958}, dislocations have been identified with torsion. Subsequent developments in metric-affine formulations associated disclinations with curvature and point defects with traces of non-metricity \cite{Dereli1987}-\cite{roychowdhury2017}. 
In this framework, the underlying geometry is typically taken to be Riemann--Cartan or Riemann--Cartan--Weyl.

However, this identification leads to a structural tension. In Riemann--Cartan geometry the full curvature decomposes as $R^a{}_b = R^a{}_b(\gamma) + D(\gamma)K^a{}_b + K^a{}_c \wedge K^c{}_b$, so that curvature already contains torsion-dependent contributions. Here $\gamma$ represents the Levi-Civita connection, $D(\gamma)$ denotes the covariant exterior derivative with respect to $\gamma$ and $K^a{}_b$ is the contortion 1-form. Associating disclinations with the full curvature therefore produces a hierarchy in which one geometric variable implicitly contains another, leading to a structural overcounting in the defect classification. By contrast, classical defect theory without differential geometry predicts the opposite hierarchy: a dislocation characterized by a Burgers vector may exist independently, and the introduction of a disclination modifies it \cite{dewit1970i}-\cite{dewit1973iv}. This mismatch indicates that the geometric organization of defect variables requires refinement.

A second structural issue concerns the absence of a genuine metric formulation. In metric-affine geometry the orthonormal coframe and the affine connection constitute independent dynamical variables, whereas in Riemannian geometry the Levi-Civita connection is completely determined by the metric. A geometrically economical theory is expected to admit a formulation in which all dynamical quantities can ultimately be expressed in terms of metric data. Metric teleparallel, symmetric teleparallel, and general teleparallel geometries satisfy this requirement \cite{adak2023ijgmmp}, whereas generic metric-affine formulations do not.

A further issue concerns the differential order of the variational field equations. This point depends on the choice of fundamental variables. In a purely metric formulation, the curvature contains
second derivatives of the metric, and generic curvature-squared terms may therefore lead to fourth-order metric field equations. In a first-order metric-affine formulation, where the affine connection is treated as an independent variable, curvature is
first order in the connection and such a conclusion does not follow automatically. Thus, the motivation adopted here is not a universal Ostrogradsky obstruction associated with curvature, but rather the
availability of a curvature-free formulation in which quadratic torsion and non-metricity invariants can be constructed from first-derivative geometric quantities and can lead to second-order field equations in the corresponding metric formulation.

These considerations motivate a reformulation of defect geometry within general teleparallel geometry, characterized by vanishing total curvature but nontrivial torsion and non-metricity. The curvature-free formulation adopted here reorganizes the local defect variables in terms of selected torsional and non-metric sectors and thereby avoids the particular bookkeeping problem that arises when a curvature tensor containing torsion-dependent contributions is itself assigned to an independent defect species. This point should not be confused with the intrinsic origin dependence of the translational part of a Euclidean or affine holonomy in the presence of a nontrivial rotational part. The latter is a group-theoretic property and is not removed merely by imposing vanishing local curvature. Quadratic Lagrangians constructed from torsion and
non-metricity yield second-order field equations \cite{adak2023ijgmmp}. Moreover, teleparallel geometry admits a metric formulation.

In the present work we develop a generalized teleparallel geometric theory of defects. A reduced set of effective defect variables is constructed from selected irreducible sectors of torsion and non-metricity in a curvature-free framework. The torsion trace is retained as an effective Burgers-type variable, the vectorial non-metricity sector $P$ as an effective Frank-type variable, and the non-metricity trace $Q$ as a point-defect variable, while the scalar
torsional sector is retained as an additional effective degree of freedom. This construction provides a reduced local geometric classification of defect variables.

\section{Metric affine geometry}

Metric-affine geometry is defined by three variables; $(M,g,\nabla)$, where $M$ is a three-dimensional differentiable and orientable manifold, $g$ is a symmetric non-degenerate covariant metric tensor, and $\nabla$ is an affine connection. A manifold can essentially be regarded as a set of points. In order to label these points, a coordinate system such as $x^a$, $a=1,2,3$, is introduced. In this way, the coordinate basis covectors (1-forms) $dx^a$ and their dual coordinate basis vectors $\partial/\partial x^a$ are constructed in a natural manner. The metric tensor is then written as $g=g_{ab} dx^a \otimes dx^b$, and the affine connection is determined by the affine connection 1-form $\widehat{\omega}^a{}_b$ via $\nabla \partial/\partial x^a := \partial/\partial x^b \otimes \widehat{\omega}^b{}_a$ and $\nabla dx^a := - \widehat{\omega}^a{}_b \wedge dx^b$, where $g_{ab}=g_{ba}$ are the covariant components of the metric in the coordinate basis, the symbol $\otimes$ denotes the tensor product, and the symbol $\wedge$ denotes the exterior product. Accordingly, the covariant derivative of a vector $\vec{U} = \partial/\partial x^a \otimes \widehat{U}^a$ and of its dual covector $\tilde{U} = \widehat{U}_a \wedge dx^a$ are given as follows
  \begin{align}
      \nabla \vec{U} := \partial/\partial x^a \otimes D \widehat{U}^a \qquad \text{and} \qquad \nabla \tilde{U} := D \widehat{U}_a \wedge dx^a.
  \end{align}
Here $D$ denotes the covariant exterior derivative, and
  \begin{align}
      D \widehat{U}^a = d \widehat{U}^a + \widehat{\omega}^a{}_b \wedge \widehat{U}^b \qquad \text{and} \qquad D \widehat{U}_a = d \widehat{U}_a - \widehat{\omega}^b{}_a \wedge \widehat{U}_b ,
  \end{align}
where $d$ is the exterior derivative. In this context, the affine connection $\nabla$ is sometimes also referred to as the covariant derivative. The covariant component $\widehat{U}_a$ of a vector and the contravariant component $\widehat{U}^a$ of a covector are 0-forms. For notational simplicity, when multiplying by a 0-form, the symbol $\wedge$ is not written explicitly.

The naturally arising coordinate basis covectors $dx^a$ can be transformed into orthonormal basis covectors $e^a$ by a linear transformation, $e^a = h^a{}_b dx^b$, where $h^a{}_b$ is known as the triad (dreibein). In fact, since the metric takes the form $g = \delta_{ab} e^a \otimes e^b$ in the orthonormal basis, with $\delta_{ab}$ denoting the Kronecker delta, the full name of $e^a$ is the metric-orthonormal basis 1-form, but we refer to it shortly as the orthonormal basis 1-form (or orthonormal coframe). Accordingly, we list below the fundamental relations for the transition from the coordinate basis to the orthonormal basis,
  \begin{subequations}
      \begin{align}
          &e^a = h^a{}_b dx^b  \quad \Leftrightarrow \quad
          \partial_a = {h^{-1}}^b{}_a \frac{\partial}{\partial x^b} , \\ 
          &\delta_{ab} = {h^{-1}}^c{}_a {h^{-1}}^d{}_b g_{cd}  , \\
          &\omega^a{}_b = h^a{}_c \widehat{\omega}^c{}_d {h^{-1}}^d{}_b     + h^a{}_c d{h^{-1}}^c{}_b .
      \end{align}
  \end{subequations}
Here the relations $h^a{}_c {h^{-1}}^c{}_b = \delta^a_b$ and $ {h^{-1}}^a{}_c h^c{}_b = \delta^a_b$ hold. It is worth noting that, in the above, tensors transform homogeneously, whereas the full connection 1-form transforms inhomogeneously, showing its non-tensorial aspect. In addition, as can be seen, when setting up the notation we distinguish coordinate components from orthonormal components not through the indices but through the root character; namely, either by writing different root characters as in $g_{ab}$ and $\delta_{ab}$, or by placing a ``widehat'' symbol on the root character as in $\omega^a{}_b$ and $\widehat{\omega}^c{}_d$. However, we shall carry out almost all discussions and formulae in the orthonormal basis, which is independent of the choice of coordinate system by definition.

An equivalent and often more geometrically transparent way to characterize metric-affine geometry is to employ the triple $(Q_{ab}, T^a , R^a{}_b)$. These objects represent, respectively, the non-metricity 1-form, the torsion 2-form, and the full curvature 2-form, all of which are defined via the Cartan structure equations,
 \begin{subequations}\label{eq:cartan-ort}
 \begin{align}
     Q_{ab} &:= -\frac{1}{2} D\delta_{ab}  = \omega_{(ab)} , \label{eq:nonmetric}\\
     T^a &:= De^a = de^a + \omega^a{}_b \wedge e^b, \label{eq:tors}\\
     R^a{}_b &:= D\omega^a{}_b := d \omega^a{}_b + \omega^a{}_c \wedge \omega^c{}_b. \label{eq:curv}
 \end{align}
 \end{subequations}
In the first equation, we have used the identity $d\delta_{ab}=0$, which follows from the fact that the orthonormal covariant components of the metric $\delta_{ab}$ consist solely of the constant values $0$ and $1$. The covariant exterior derivative of the affine connection 1-form is not well defined. For this reason, the notation $D\omega^a{}_b$ is introduced together with the symbol $:=$, emphasizing that it serves as a convenient formal expression motivated by analogy rather than a genuine tensorial operation.  

Throughout this work, symmetrization and antisymmetrization over indices are denoted, in accordance with standard convention, by parentheses $( \cdot)$ and brackets $[\vdot]$, respectively. Explicitly, $\omega_{(ab)} = \frac{1}{2} (\omega_{ab} + \omega_{ba})$ and $\omega_{[ab]} = \frac{1}{2} (\omega_{ab} - \omega_{ba})$. The tensor-valued differential forms defined above are not entirely independent; instead, they satisfy three Bianchi identities, which encode the underlying geometric consistency conditions of the metric-affine structure,
  \begin{align} \label{eq:bianchi-identities}
    DR^a{}_b = 0,   \qquad DT^a = R^a{}_b \wedge e^b , \qquad  DQ_{ab} = R_{(ab)}.
  \end{align}
For later use and for comparison with more traditional tensorial formulations, it is useful to recall that one may always translate between the differential-form language and component-based tensor notation via the relations
  \begin{align}
      Q_{ab}=Q_{abc} e^c , \qquad T^a = \frac{1}{2} T^a{}_{bc} e^b \wedge e^c , \qquad R^a{}_b = \frac{1}{2} R^a{}_{bcd} e^c \wedge e^d .
  \end{align}
Several immediate structural consequences follow from these definitions. The $(0,3)$-type non-metricity tensor $Q_{abc}$ is symmetric in its first two indices, while the $(1,2)$-type torsion tensor $T^a{}_{bc}$ is antisymmetric in its last two indices. Similarly, the $(1,3)$-type full curvature tensor $R^a{}_{bcd}$ is antisymmetric in its last two indices. Moreover, as a direct consequence of the first Bianchi identity, the condition $Q_{ab}=0$ implies that the curvature tensor $R_{abcd}$ also becomes antisymmetric in its first pair of indices, recovering the familiar symmetry properties of Riemannian geometry.  

Finally, for practical calculations in an orthonormal frame, it is advantageous to keep in mind the following identities involving the exterior derivative and the covariant exterior derivative of the metric components,
   \begin{subequations}
  \begin{align}
      d\delta_{ab}&=0, & d\delta^{ab}&=0, & d\delta^a_b&=0 , \\
      D\delta_{ab} &= -2Q_{ab}, & D\delta^{ab}&=+2Q^{ab}, & D\delta^a_b&=0 .
  \end{align}
   \end{subequations}
Accordingly, in an orthonormal frame, indices may be raised or lowered freely when acting with the exterior derivative $d$, whereas particular care must be exercised when performing the same operations in expressions involving the covariant exterior derivative $D$.

The full connection 1-form can be decomposed into its Riemannian content and non-Riemannian content as follows \cite{adak2023ijgmmp}-\cite{hehl1995} 
   \begin{align} \label{eq:connec-decom}
     \omega_{ab}=\gamma_{ab} + L_{ab} .
 \end{align}
Here, the Riemannian content $\gamma_{ab} = -\gamma_{ba}$ is the Levi-Civita connection 1-form, defined by $\gamma^a{}_b \wedge e^b = -de^a$, whereas the non-Riemannian content $L_{ab}$ is known as the distortion 1-form in alternative theories of gravity. However, since we believe that it reflects the essence of the problem more accurately, we adopt the terminology defect 1-form in the context of the general teleparallel geometric theory of defects
 \begin{align} \label{eq:kusur-formu}
     L_{ab} =  \underbrace{ \frac{1}{2} \left[ \iota_a T_b - \iota_b T_a - (\iota_a \iota_b T_c) e^c \right]}_{contortion} + \underbrace{ ( \imath_b Q_{ac} - \imath_a Q_{bc} ) e^c + Q_{ab} }_{disformation} .
 \end{align}
where $\iota_a \equiv \iota_{X_a}$ denotes the interior product of the exterior algebra, satisfying $\iota_{X_b} e^a = \delta^a_b$. In the literature, instead of the torsion 2-form, the contortion 1-form $K_{ab}=-K_{ba}$ is sometimes used, defined via $K^a{}_b \wedge e^b = T^a$. If the non-metricity vanishes, the full connection is referred to as a metric-compatible connection, and if, in addition, the torsion also vanishes, it reduces to the Levi-Civita connection. Accordingly, the full curvature 2-form $R^a{}_b = R^a{}_b(\omega)$ given in (\ref{eq:curv}) can be decomposed into its Riemannian and non-Riemannian parts as
   \begin{align} \label{eq:decompos-curv}
     R^a{}_b = R^a{}_b(\gamma) + D(\gamma) L^a{}_b + L^a{}_c \wedge L^c{}_b
 \end{align}
where $R^a{}_b(\gamma)$ is the Riemannian curvature 2-form and $D(\gamma) L^a{}_b$ denotes the covariant exterior derivative of the defect 1-form with respect to the Levi-Civita connection
 \begin{subequations}
     \begin{align}
         R^a{}_b (\gamma) &= d\gamma^a{}_{b} + \gamma^a{}_{c} \wedge \gamma^c{}_{b},  \\
         D(\gamma)L^a{}_b &= dL^a{}_b +  \gamma^a{}_{c} \wedge L^c{}_b - \gamma^c{}_{b} \wedge L^a{}_c . \label{eq:tildaD}
     \end{align}
 \end{subequations}
Affine geometries are classified according to whether the non-metricity, torsion, and/or full curvature tensors vanish or not. In this classification, a geometry characterized by $Q_{ab} \neq 0$, $T^a \neq 0$, and $R^a{}_b =0$ is referred to as general teleparallel geometry, and it admits a formulation as a metric geometry \cite{adak2023ijgmmp}.

The torsion tensor, whose last two indices are antisymmetric, has nine independent components
  \begin{align}
      T_{abc} = \{ T_{112}, T_{113}, T_{123}, T_{212}, T_{213}, T_{223}, T_{312}, T_{313}, T_{323} \} .
  \end{align}
Accordingly, the torsion 2-form can be decomposed into the following irreducible pieces
   \begin{subequations}\label{decomp_torsion}
  \begin{align}
      &T^a = \overset{(1)}{T^a} + \overset{(2)}{T^a} + \overset{(3)}{T^a} \qquad \text{where} \\
      &\overset{(2)}{T^a} = \frac{1}{2} e^a \wedge T , \qquad \overset{(3)}{T^a} = \frac{1}{3} \iota^a S , \qquad \overset{(1)}{T^a} =  T^a - \overset{(2)}{T^a} - \overset{(3)}{T^a} .
  \end{align}
   \end{subequations}
Here, $T$ denotes the torsion trace 1-form and $S$ denotes the torsion scalar 3-form, and the first piece $\overset{(1)}{T^a}$ satisfies the following properties
      \begin{align}
          T = \iota_a T^a, \qquad  S = e_a \wedge T^a , \qquad 
       \iota_a \overset{(1)}{T^a} =0, \qquad e_a \wedge \overset{(1)}{T^a} =0 .
  \end{align}
The physical interpretation of the irreducible torsion pieces requires some care. In the geometric theory of distributed dislocations, torsion is naturally associated with the local dislocation density rather than directly with a single Burgers vector. In order to make this correspondence explicit, we introduce the second-rank dislocation-density tensor $\alpha^{ab}$ by Hodge dualizing the vector-valued torsion 2-form,
\begin{equation}
    *T^{a}=\alpha^{ab}e_b,
    \qquad
    \alpha^{ab}
    =
    \frac{1}{2}\epsilon^{bcd}T^{a}{}_{cd}.
\label{dislocation_density}
\end{equation}
Here $\alpha^{ab}$ is the local dislocation-density tensor associated with the torsion field. Notice that different sign or transpose conventions for Nye-type dislocation-density tensors are used in the literature. Throughout the present work, our convention is fixed by (\ref{dislocation_density}).

It is important to distinguish the local dislocation density from the Burgers vector itself. The latter is a cumulative measure associated with a closed material circuit and can be defined geometrically through the closure failure of its affine development \cite{Ozakin_Yavari_2014}. Let 
$C:[0,\ell]\rightarrow M$ be a closed curve based at 
$p=C(0)=C(\ell)$, and let $P_{C(t)\rightarrow p}$ denote parallel transport from $C(t)$ to $p$ with respect to the affine connection. The affine development $\dot{c}(t)\in T_pM$ is defined by
\begin{equation}
    \dot{c}(t)
    =
    P_{C(t)\rightarrow p}\dot{C}(t),
    \qquad
    c(0)=0 .
\label{affine_development}
\end{equation}
The Burgers vector associated with the circuit $C$ is then the closure failure
\begin{equation}
    B(p;C)
    :=
    c(\ell)-c(0)
    =
    \int_{0}^{\ell}
    P_{C(t)\rightarrow p}\dot{C}(t)\,dt .
\label{burgers_affine}
\end{equation}
For infinitesimal closed circuits, the areal density of this closure 
failure is governed by the torsion tensor. Thus, $T^{a}$, or equivalently $\alpha^{ab}$, represents the local dislocation density, whereas $B(p;C)$ represents the Burgers vector associated with a finite closed 
circuit.

The irreducible decomposition of torsion given in (\ref{decomp_torsion}) has a direct interpretation in terms of the irreducible decomposition of the dislocation-density tensor. In three dimensions, any second-rank tensor $\alpha_{ab}$ can be decomposed as
\begin{equation}
    \alpha_{ab}
    =
    \alpha_{\langle ab\rangle}
    +
    \alpha_{[ab]}
    +
    \frac{1}{3}
    (\mathrm{tr}\,\alpha)\delta_{ab},
\label{alpha_decomposition}
\end{equation}
where
\begin{equation}
    \alpha_{\langle ab\rangle}
    :=
    \alpha_{(ab)}
    -
    \frac{1}{3}
    (\mathrm{tr}\,\alpha)\delta_{ab}
\end{equation}
denotes the symmetric traceless part. Through (\ref{dislocation_density}), the three irreducible torsion sectors correspond precisely to these three pieces:
\begin{equation}
    \overset{(1)}{T^{a}}
    =
    \alpha^{\langle ab\rangle}*e_{b},
    \qquad
    \overset{(2)}{T^{a}}
    =
    \alpha^{[ab]}*e_{b},
    \qquad
    \overset{(3)}{T^{a}}
    =
    \frac{1}{3}
    (\mathrm{tr}\,\alpha)*e^{a}.
\label{torsion_alpha_decomposition}
\end{equation}
Accordingly, the three irreducible pieces contain $5$, $3$, and $1$ independent components, respectively. In particular, the torsion trace 1-form $T=\iota_a T^a$ is dual to the antisymmetric part of the dislocation-density tensor, whereas the torsion scalar 3-form satisfies
\begin{equation}
    S=e_a\wedge T^a = (\mathrm{tr}\,\alpha)*1.
\label{S_alpha_trace}
\end{equation}
The remaining five-component field $\overset{(1)}{T^a}$ corresponds to the symmetric traceless part of the dislocation-density tensor and therefore also carries genuine dislocation information. In the present work, however, we deliberately restrict the subsequent analysis to the vectorial and scalar torsional sectors. The omission of $\overset{(1)}{T^a}$ should therefore be understood as a truncation defining
the reduced model considered here, rather than as a statement that this sector is physically meaningless.

The physical meaning of these irreducible sectors becomes particularly transparent for a locally straight dislocation. Let $\vec{\xi}$ denote the unit tangent vector to the dislocation line, $\vec{b}$ its Burgers vector, and $\varrho_{\ell}$ a local line-density factor. The corresponding dislocation-density tensor can be written as
\begin{equation}
    \alpha_{ab} = \varrho_{\ell} b_a \xi_b .
\label{straight_dislocation_density}
\end{equation}
Using Eqs.~(\ref{alpha_decomposition}) and (\ref{torsion_alpha_decomposition}), one obtains
\begin{equation}
    T^{\sharp}
    =
    \varrho_{\ell}
    \left(
    \vec{\xi}\times\vec{b}
    \right),
    \qquad
    S
    =
    \varrho_{\ell}
    \left(
    \vec{b}\cdot\vec{\xi}
    \right)*1 ,
\label{T_S_dislocation}
\end{equation}
where $\sharp$ denotes the musical isomorphism mapping the 1-form $T$ 
to its corresponding vector. Moreover, the symmetric traceless part is
\begin{equation}
    \alpha_{\langle ab\rangle}
    =
    \frac{\varrho_{\ell}}{2}
    \left(
    b_a\xi_b+\xi_a b_b
    \right)
    -
    \frac{\varrho_{\ell}}{3}
    \left(
    \vec{b}\cdot\vec{\xi}
    \right)\delta_{ab}.
\label{sym_traceless_dislocation}
\end{equation}
These relations show explicitly why the torsion trace $T$ alone cannot 
be identified with a general Burgers vector. For a screw dislocation, 
$\vec{b}\parallel\vec{\xi}$, and therefore $T=0$, while $S$ and 
$\overset{(1)}{T^a}$ are in general non-vanishing. For an edge dislocation, 
$\vec{b}\perp\vec{\xi}$, and hence $S=0$, while both 
$\overset{(1)}{T^a}$ and $\overset{(2)}{T^a}$ are in general non-vanishing. Thus, a completely general representation of conventional dislocation
distributions requires the full vector-valued torsion 2-form $T^a$,
rather than only its trace sector. The reduced model considered below
retains only the vectorial and scalar projections $T$ and $S$; hence, configurations whose complete local description requires the symmetric-traceless sector $\overset{(1)}{T^a}$ lie outside the scope of this truncation.

The analysis above requires a refinement of the physical interpretation of the torsional variables. The Burgers vector should not be identified directly with the torsion trace $T$. Rather, the full vector-valued torsion 2-form represents the local dislocation-density field,
\begin{equation}
    \text{local dislocation density}
    \ \sim\
    \alpha^{a}{}_{b}
    \ \sim\
    T^{a},
\label{dislocation_identification}
\end{equation}
whereas the Burgers vector is an integrated closure-failure measure.

The quantities $T$ and $S$ characterize different irreducible projections of this local dislocation-density field:
\begin{equation}
    T
    \ \sim\
    \operatorname{skew}\alpha,
    \qquad
    S
    =
    (\mathrm{tr}\,\alpha)*1,
    \qquad
    \overset{(1)}{T^a}
    \ \sim\
    \operatorname{dev\,sym}\alpha .
\label{irreducible_dislocation_identification}
\end{equation}
Thus, $T$ and $S$ should not be interpreted as complete conventional
Burgers data by themselves. In the reduced geometric model developed
below, however, they are retained as independent irreducible torsional
sectors, while $\overset{(1)}{T^a}$ is consistently set to zero as a modelling
truncation.

In particular, introducing the scalar field $\rho$ through
\begin{equation}
    S=\rho *1,
\end{equation}
one obtains
\begin{equation}
    \rho=\mathrm{tr}\,\alpha .
\label{rho_trace_alpha}
\end{equation}
Therefore, $\rho$ parametrizes the scalar trace sector of the local dislocation-density tensor. In the present reduced theory it is treated as an additional scalar torsional degree of freedom, rather than as a
new conventional defect species unrelated to dislocations.

Accordingly, the present work focuses on the reduced torsional sector obtained by imposing $\overset{(1)}{T^a}=0$. This assumption restricts the class of conventional dislocation-density fields described by the model but preserves the vectorial and scalar irreducible sectors that will be used as effective defect variables below. The torsion 2-form therefore reduces to
   \begin{align} \label{eq:burulma-ayris1}
      T^a = \frac{1}{2} e^a \wedge T +  \frac{1}{3} \iota^a S .
  \end{align}

The non-metricity tensor, whose first two indices are symmetric, has eighteen independent components
  \begin{align}
      Q_{abc} &= \{ Q_{111}, Q_{112}, Q_{113}, Q_{121}, Q_{122}, Q_{123}, Q_{131}, Q_{132}, Q_{133}, \nonumber \\
     & \qquad Q_{221}, Q_{222}, Q_{223}, Q_{231}, Q_{232}, Q_{233}, Q_{331}, Q_{332}, Q_{333} \} .
  \end{align}
The non-metricity 1-form can be decomposed as follows
\begin{subequations}
  \begin{align}
      &Q_{ab} = \overset{(1)}{Q_{ab}} + \overset{(2)}{Q_{ab}} + \overset{(3)}{Q_{ab}} + \overset{(4)}{Q_{ab}} \qquad \text{where} \\   
     &\overset{(2)}{Q_{ab}} =  -\frac{1}{3} \left( \iota_a \widehat{N}_b + \iota_b \widehat{N}_a \right) , \\
     &\overset{(3)}{Q_{ab}} = \frac{3}{10}  \left[ (\iota_a P) e_b + (\iota_b P) e_a - \frac{2}{3} \delta_{ab} P \right] , \\
     &\overset{(4)}{Q_{ab}} =  \frac{1}{3} \delta_{ab} Q , \\
      &\overset{(1)}{Q_{ab}} = Q_{ab} - \overset{(2)}{Q_{ab}} - \overset{(3)}{Q_{ab}} - \overset{(4)}{Q_{ab}} .
  \end{align}
   \end{subequations}
In the above, the decomposition $Q_{ab} = \overline{Q}_{ab} + \frac{1}{3} \delta_{ab} Q$ has been used, together with the following definitions
      \begin{align}\label{independent_parts_Qab}
       Q=\delta^{ab}Q_{ab} , \qquad  N_a = \overline{Q}_{ab} \wedge e^b , \qquad P = \left( \iota^a \overline{Q}_{ab} \right) e^b, \qquad \widehat{N}_a = N_a-\frac{1}{2}e_a\wedge P .
      \end{align}
The quantities $N_a$ and $P$ are not independent. Indeed, the definitions in (\ref{independent_parts_Qab}), together with
$\delta^{ab}\overline{Q}_{ab}=0$, imply
\begin{equation}
\iota^a N_a=P,
\qquad
e^a\wedge N_a=0.
\end{equation}
Consequently, the combination $\widehat{N}_a$ satisfies
\begin{equation}
\iota^a\widehat{N}_a=0,
\qquad
e^a\wedge\widehat{N}_a=0.
\end{equation}
In the literature, $Q$ is known as the non-metricity trace 1-form. Accordingly, the following properties hold
      \begin{align}
      \delta^{ab} \overline{Q}_{ab}=0, \qquad \delta^{ab} \overset{(1)}{Q_{ab}} = 0 , \qquad \iota^a \overset{(1)}{Q_{ab}} = 0 , \qquad  e^a \wedge \overset{(1)}{Q_{ab}} = 0 .
  \end{align}
Moreover, the second and third pieces are separated such that
\begin{align}
 \overset{(2)}{Q_{ab}}\wedge e^b=\widehat{N}_a,
\qquad
(\iota^a \overset{(3)}{Q_{ab}})e^b=P,
\qquad
\overset{(3)}{Q_{ab}}\wedge e^b
=
\frac{1}{2}e_a\wedge P.   
\end{align}
The component content of this decomposition can now be counted without redundancy. The 1-forms $Q$ and $P$ each contain three independent components, and therefore $\overset{(4)}{Q_{ab}}$ and $\overset{(3)}{Q_{ab}}$ contain three components each. Although $N_a$ is a vector-valued 2-form with nine raw components, its $P$-dependent part has already been separated in (\ref{independent_parts_Qab}). The remaining quantity $\widehat{N}_a$ obeys the four
independent conditions $\iota^a\widehat{N}_a=0$ and
$e^a\wedge\widehat{N}_a=0$, and hence contains five independent components. Therefore $\overset{(2)}{Q_{ab}}$ has five independent components. The remaining irreducible piece $\overset{(1)}{Q_{ab}}$ consequently contains seven independent components. Thus, the eighteen independent components of
non-metricity $Q_{ab} = \overset{(1)}{Q_{ab}} + \overset{(2)}{Q_{ab}} + \overset{(3)}{Q_{ab}} + \overset{(4)}{Q_{ab}} \qquad$ are decomposed as $18=7+5+3+3$. Hence, within the reduced general teleparallel geometric theory of defects developed here, the following effective identifications
are adopted:
\begin{subequations}
\begin{align}
P
&\sim
\text{effective Frank-type covector}
\sim
\text{disclination sector},
\\
Q
&\sim
\text{point-defect covector}
\sim
\text{point-defect sector}.
\end{align}
\end{subequations}
Here the symbol "$\sim$" denotes an association between an irreducible geometric sector and the effective local variable used to parametrize the corresponding defect sector; it should not be
understood as equality with the conventional integrated Frank measure. In the present work we deliberately restrict the subsequent analysis to the vector-trace sector parametrized by $P$ and $Q$.
Accordingly, we set $\overset{(1)}Q_{ab}=0,
\overset{(2)}Q_{ab}=0$. This restriction defines the reduced sector studied here and
should not be interpreted as implying that the omitted
seven- and five-component sectors are physically meaningless. Since$\overset{(2)}Q_{ab}\wedge e^b=\widehat{N}_a$, the condition $\overset{(2)}Q_{ab}=0$ implies
$\widehat{N}_a=0$, or equivalently $N_a=\frac{1}{2}e_a\wedge P$. The reduced non-metricity 1-form suitable for use in the general teleparallel geometric theory of defects takes the form
  \begin{align} \label{eq:nonmetricity-ayris1}
     Q_{ab} =   \frac{3}{10}  \left[ (\iota_a P) e_b + (\iota_b P) e_a - \frac{2}{3} \delta_{ab} P \right] +  \frac{1}{3} \delta_{ab} Q .
  \end{align}

In summary, there are three well-known types of crystal defects: point defects (extra-matter), dislocations, and disclinations. In contrast, within the general teleparallel geometric framework developed in this work, there are four effective geometric sectors: the torsion trace 1-form $T$ derived from torsion, the scalar 3-form $S$ derived from torsion, the 1-form $P$ derived from non-metricity, and the 1-form $Q$ derived from non-metricity. These sectors contain, respectively, three, one, three, and three independent components. Their interpretation as defect variables will be specified more precisely in Sec.~5.

\section{Elastic deformations} \label{sec:elastic-deformations}

\begin{figure}[ht]
  \centering 	
\includegraphics[width=0.8\textwidth, angle=0]{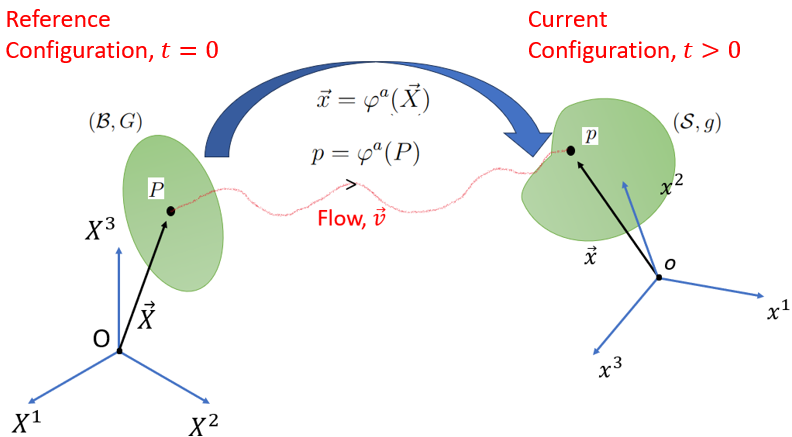}
  \caption{Continuum mechanics. $\mathcal{B}$ is the body manifold, $\mathcal{S}$ is the spatial manifold.}
  \label{fig:surekli_ortam}
\end{figure}

Under physical effects such as forces, temperature, and similar influences, a part (region) or the entirety of a continuum may undergo translational motion, rotational motion, and/or deformation. The branch of science that investigates all of these phenomena is called continuum mechanics. In general, continuum mechanics can be summarized by the schematic illustration shown in Figure~\ref{fig:surekli_ortam}.

At time $t=0$, we denote the material or reference manifold of the body by $(\mathcal{B},G)$. For a defect-free elastic body this manifold may be identified with an undeformed stress-free reference
configuration. In the presence of distributed defects, however, $\mathcal{B}$ should more generally be regarded as an intrinsic material manifold whose geometry describes the locally relaxed structure of the body. Such a material geometry need not admit a
globally stress-free realization in Euclidean space. In the mechanical engineering literature, the coordinates of points on the body are denoted by capital letters; a point is written as $P=\vec{X}$ or $P=X^A=(X,Y,Z)$, with $A=1,2,3$. These coordinates are also called Lagrangian coordinates. Accordingly, the metric used to measure the distance between two infinitesimally close points in $\mathcal{B}$ is given by $dS^2 := G = G_{AB} \, dX^A \otimes dX^B$. Thus, for the Riemannian geometry $(\mathcal{B},G)$, one has the coordinate frame $\partial/\partial X^A$ and its dual coordinate coframe $dX^A$. 

Naturally, by means of $G=\delta_{AB} E^A \otimes E^B$, one can pass to an orthonormal coframe $E^A$ and the corresponding orthonormal frame $\partial_A$ using the triads $H^A{}_B$ and the inverse triads ${H^{-1}}^A{}_B$,
\begin{align}
     E^A = H^A{}_B dX^B \qquad \text{and} \qquad \partial_A = {H^{-1}}^B{}_A \frac{\partial}{\partial X^B}.
\end{align}
Here, due to the duality relation $E^A(\partial_B)=\delta^A_B$, the identity $H^A{}_C {H^{-1}}^C{}_B = \delta^A_B$ holds. In addition, the Levi--Civita connection 1-form used in the parallel transport of vectors, or more generally tensors, can be computed with the aid of the orthonormal coframe via $\Gamma^A{}_B \wedge E^B = -dE^A$. Finally, the invariant volume element associated with the geometry $(\mathcal{B},G)$ is given by $* \mathbb{1} = VOL^3 = \frac{1}{3!} \epsilon_{ABC} E^A \wedge E^B \wedge E^C$. Here, the asterisk denotes the Hodge mapping, $\epsilon_{ABC}$ is the totally antisymmetric epsilon tensor, and the orientation of the manifold is fixed by $\epsilon_{123}=+1$.

At a later time $t>0$, the new state of the body is referred to as the current configuration, the deformed configuration, or the spatial manifold $\mathcal{S}$. In this configuration, mechanical engineers conventionally denote the coordinates of points on the body by lowercase letters; a point is written as $p=\vec{x}$ or $p=x^a=(x,y,z)$, with $a=1,2,3$. These are also called Eulerian coordinates. Definitions analogous to those given above can be introduced here as well. Accordingly, for the Riemannian geometry $(\mathcal{S},g)$, the metric is given by $ds^2 := g = g_{ab} \, dx^a \otimes dx^b = \delta_{ab} \, e^a \otimes e^b$, together with the coordinate frame $\partial/\partial x^a$, the coordinate coframe $dx^a$, the orthonormal frame $\partial_a$, the orthonormal coframe $e^a$, the triads $h^a{}_b$, and the inverse triads ${h^{-1}}^a{}_b$,
\begin{align}
  e^a = h^a{}_b dx^b \qquad \text{and} \qquad \partial_a = {h^{-1}}^b{}_a \frac{\partial}{\partial x^b}.
\end{align}
Here, due to the duality relation $e^a(\partial_b)=\delta^a_b$, the identity $h^a{}_c {h^{-1}}^c{}_b = \delta^a_b$ holds. The Levi--Civita connection 1-form can be computed with the aid of the orthonormal coframe via $\gamma^a{}_b \wedge e^b = -de^a$. Likewise, for the configuration $(\mathcal{S},g)$, the invariant volume element is given by $* 1 = vol^3 = \frac{1}{3!} \epsilon_{abc} e^a \wedge e^b \wedge e^c $.

In standard continuum mechanics, the material (Lagrangian) description is formulated on $(\mathcal B,G)$ with capital-letter coordinates, whereas the spatial (Eulerian) description is formulated on $(\mathcal S,g)$ with lowercase coordinates. In this setting, the spatial manifold $\mathcal{S}$ may be regarded as a region that deforms and evolves within a larger continuum, represented by a manifold $\mathcal{M}$, in which case $\partial \mathcal{S}$ denotes the boundary of this region.\footnote{\label{ft:deniz-poset} One may visualize $\mathcal{M}$ as a flowing river and $\mathcal{S}$ as a portion of this river (enclosed by an imaginary, transparent, very thin membrane), with $\partial \mathcal{S}$ representing this membrane, i.e.\ a closed surface.} As long as this process takes place within the elastic regime, the geometry of $\partial \mathcal{S}$ may change while the total mass contained inside remains the same. Thus, one of the fundamental equations of continuum mechanics is the mass conservation equation \cite{frankel2012}, \cite{reddy2013},
\begin{align}
    \mathcal{L}_{\vec{v} + \partial/\partial t} m =0 \qquad \Rightarrow \qquad \frac{\partial \rho_m}{\partial t} + \Vec{\nabla} \vdot (\rho_m \Vec{v}) =0 .
\end{align}
Here $\rho_m$ is the mass density scalar, $m=\rho_m *1$ is the mass 3-form, and $\mathcal{L}$ denotes the Lie derivative. Another fundamental equation of continuum mechanics is the Cauchy equation of motion of $\mathcal{S}$ (the so-called balance of linear momentum) \cite{frankel2012}, \cite{reddy2013}, 
\begin{align}
     m \left( \frac{\partial v^a} {\partial t} + \iota_{\Vec{v}} Dv^a  \right) = m f^a_{bd} +  D\tau^a \label{eq:cauchy-hareket-denk2}.
\end{align}

In (\ref{eq:cauchy-hareket-denk2}), $D$ denotes the covariant exterior derivative associated with the Levi--Civita connection of the spatial Riemannian manifold $(\mathcal{S},g)$. Here the vector $\vec{f}_{bd}$ represents the body force per unit mass acting at every point in the interior and on the boundary of $\mathcal{S}$, such as gravitational and electromagnetic forces, while the term $\tau = \partial_a \otimes \tau^a$ denotes the vector-valued Cauchy stress 2-form representing the surface forces acting on the closed surface $\partial \mathcal{S}$. The latter can be written as $\tau^a = \sigma^{ab} *e_b$, where $\sigma^{ab}$ are the components of the Cauchy stress tensor, i.e.\ scalar (0-form) fields. Finally, for a vector field $\vec{U}$ and a differential form $\alpha$, the Lie derivatives with respect to a vector field $\vec{V}$ are given by
\begin{align}
    \mathcal{L}_{\vec{V}}\vec{U} = [\vec{V},\vec{U}]  \qquad \text{and} \qquad \mathcal{L}_{\vec{V}} \alpha = \iota_{\vec{V}} d\alpha + d\iota_{\vec{V}} \alpha .
\end{align}

For finite deformations, the kinematics should be formulated before specifying a constitutive law. Let $\varphi:\mathcal{B}\longrightarrow\mathcal{S}$
denote a regular configuration of the body, where
$(\mathcal{B},G)$ is the material manifold and
$(\mathcal{S},g)$ is the spatial manifold. In local coordinates, the configuration is written as $x^a=\varphi^a(X^A)$. The deformation gradient is the tangent map $F:=T\varphi:
T_X\mathcal{B}
\longrightarrow
T_{\varphi(X)}\mathcal{S}$,
whose coordinate components are
\begin{equation}
\widehat F^a{}_A
=
\frac{\partial\varphi^a}{\partial X^A}.
\end{equation}
Accordingly, the pull-back of the spatial coordinate coframe is
\begin{equation}
\varphi^*(dx^a)
=
\widehat F^a{}_A\,dX^A.
\end{equation}
Using the triads introduced above, the same relation can be written in the orthonormal frames as
\begin{equation}
\varphi^*(e^a)
=
F^a{}_A E^A,
\qquad
F^a{}_A
=
h^a{}_b
\widehat F^b{}_B
H^{-1B}{}_A .
\end{equation}
For a regular configuration the inverse deformation gradient exists locally and satisfies 
$F^A{}_aF^a{}_B=\delta^A_B$ and
$F^a{}_AF^A{}_b=\delta^a_b$.
The finite elastic deformation can be characterized geometrically by comparing the material metric $G$ with the pull-back of the spatial metric. The right Cauchy--Green deformation tensor is
\begin{equation}
C
:=
\varphi^*g,
\qquad
C_{AB}
=
(g_{ab}\circ\varphi)
\widehat F^a{}_A
\widehat F^b{}_B .
\end{equation}
The corresponding Green--Lagrange strain tensor is
\begin{equation}
E
=
\frac{1}{2}
\left(
C-G
\right)
=
\frac{1}{2}
\left(
\varphi^*g-G
\right).
\end{equation}
Since in the present work we primarily employ the spatial description, it is also useful to introduce the spatial strain measure. For a regular configuration, let $c:= (\varphi^{-1})^*G = \varphi_*G$ denote the material metric transported to the current configuration. The Euler--Almansi strain tensor is then
\begin{equation}
e
=
\frac{1}{2}
\left(
g-c
\right)
=
e_{ab}e^a\otimes e^b,
\end{equation}
with
\begin{equation}
e_{ab}
=
\frac{1}{2}
\left(
\delta_{ab}
-
\delta_{AB}F^A{}_aF^B{}_b
\right)
\end{equation}
in orthonormal frames. For finite elastic deformations, the constitutive response is in
general nonlinear. For a hyperelastic material it is specified by a strain-energy density that may be written in material form as $W=W(X,G,C)$, or equivalently as $W=W(F,G,g)$. For a frame-indifferent hyperelastic material, the constitutive dependence may be expressed in terms of the right Cauchy--Green tensor $C$. For an isotropic material, isotropy further reduces this dependence to its principal invariants, $W=W(I_1,I_2,I_3)$. The second Piola--Kirchhoff stress is obtained from
\begin{equation}
S^{AB}
=
2\frac{\partial W}{\partial C_{AB}},
\end{equation}
and the corresponding Cauchy stress is
\begin{equation}
\sigma^{ab}
=
\frac{1}{J}
F^a{}_A F^b{}_B S^{AB},
\end{equation}
where $J$ denotes the local volume ratio of the deformation. The familiar Hooke law is recovered only after linearizing the finite-deformation theory about a stress-free configuration. For an isotropic material, the infinitesimal-strain limit gives
\begin{equation}
\sigma_{ab}
=
2\mu \varepsilon_{ab}
+
\lambda(\operatorname{tr}\varepsilon)\delta_{ab},
\end{equation}
where $\lambda$ and $\mu$ are the Lam\'e constants and $\varepsilon_{ab}$ denotes the infinitesimal strain obtained by linearizing the finite Euler--Almansi strain about the stress-free configuration. This relation is therefore used only as the small-strain limit of the nonlinear constitutive theory and not as the constitutive equation for general finite deformations.

The presence of distributed defects should not, in general, be identified with a loss of regularity or invertibility of the deformation map. In geometric theories of anelasticity, the configuration $\varphi:\mathcal{B}\rightarrow\mathcal{S}$ may
remain regular, while the defect content is encoded intrinsically in the geometry assigned to the material manifold. In this sense, the locally relaxed material structure may be geometrically incompatible with a global stress-free embedding in Euclidean space, and residual stresses arise when the material manifold is
embedded into the spatial manifold.

This geometric viewpoint has been developed extensively in nonlinear defect mechanics. Yavari and Goriely formulated distributed dislocations by identifying the stress-free material configuration with a flat, metric-compatible material manifold
with non-vanishing torsion, the latter representing the local dislocation density. They subsequently treated distributed disclinations using a stress-free material manifold with non-vanishing curvature and calculated the corresponding residual stress fields within nonlinear elasticity. Distributed point defects were also formulated within Weyl geometry and nonlinear elasticity \cite{Yavari_Goriely_2012_a}. The framework was later extended to combined distributions of dislocations, disclinations, and point defects, referred to as discombinations, and has also
been applied to wedge dispirations and to line and point defects in nonlinear anisotropic solids \cite{Yavari_Gorielli_2012}-\cite{Golgoon_Yavari_2018}.

The present work adopts the same general principle that distributed defect information is carried by the intrinsic material geometry rather than by a singular deformation map. Our geometrical organization of the defect variables, however, is different. Instead of assigning disclinations to a non-vanishing total
curvature, we impose $R^a{}_b=0$ and describe the reduced defect sector through selected irreducible components of torsion and non-metricity within general
teleparallel geometry. The purpose of the present section is therefore to establish the finite-deformation continuum-mechanical setting; the specific geometric identification of the effective defect variables is developed in Secs.~4 and 5.

\section{Metric affine geometric theory of defects} \label{sec:mag-theory-of-defects}

In the literature on differential–geometric approaches to topological defects in crystals, the following identifications are commonly adopted \cite{Dereli1987}–\cite{roychowdhury2017}
  \begin{subequations}
   \begin{align}
&\text{dislocation} \sim \text{Burgers vector}, \, \vec{b} \sim T^a , \\
&\text{disclination} \sim \text{general Frank vector}, \, \vec{\Omega} \sim R^a{}_b ,\\
&\text{point defect} \sim \text{no associated vector} \sim Q_{ab} .
    \end{align}
   \end{subequations}
In these models, the underlying geometric framework in which the discussion is carried out is metric–affine geometry. More precisely, since the point defect is associated with the $Q$ component of non-metricity, this geometry is referred to as Riemann–Cartan–Weyl geometry. Subsequently, physical interpretations of defects are extracted by means of the Bianchi identities (\ref{eq:bianchi-identities}). Exterior algebra in three-dimensional manifolds allows one to make the following identifications \cite{suhubi2013}: the exterior derivative of a zero-form is isomorphic to the gradient of a scalar, the exterior derivative of a one-form is isomorphic to the curl of a vector, and the exterior derivative of a two-form is isomorphic to the divergence of a vector. In light of these correspondences, one can immediately state the following results. According to the identity (\ref{eq:bianchi-identities}i), disclination lines are, analogous to magnetic field lines in the law of no magnetic monopoles, closed curves. According to the identity (\ref{eq:bianchi-identities}ii), dislocation lines, analogous to electric field lines in Gauss’s law, originate and terminate on disclinations. Finally, according to the identity (\ref{eq:bianchi-identities}iii), point-defect lines, analogous to magnetic field lines in Ampère’s law, form closed curves encircling disclination lines. 

Although the first two interpretations above are stated explicitly in the literature, the third interpretation is not articulated with the same level of clarity \cite{miri2002}. Here, we make this interpretation explicit. Moreover, by inspecting the identity (\ref{eq:bianchi-identities}ii), one often encounters the statement that curvature is the source of torsion \cite{miri2002}. Such a statement should be interpreted with care. Because the decomposition of the full curvature given in (\ref{eq:decompos-curv}) itself contains torsion-dependent contributions through the contortion, the second Bianchi identity is more appropriately regarded as a geometric compatibility relation rather than as a one-way source equation in which curvature generates torsion.

It is important to distinguish this algebraic issue from a different and intrinsic ambiguity that occurs in the holonomy description of combined dislocations and disclinations. Let a Euclidean holonomy be represented by $(\Lambda,t)$, where $\Lambda$ is its rotational part and $t$ its translational part. Under a change of origin $x=y+a$, the translational part transforms as
\begin{equation}
    t \longmapsto t' = t+(\Lambda-I)a .
\end{equation}
Therefore, whenever $\Lambda\neq I$, the translational part depends on the choice of origin. This is an intrinsic group-theoretic property of the Euclidean holonomy and is distinct from the algebraic overlap associated with the curvature decomposition in (\ref{eq:decompos-curv}). Accordingly, the curvature-free construction considered below is not claimed to remove this general holonomy property. Rather, our aim is to reorganize the local defect fields within a reduced general teleparallel sector in terms of independent effective torsional and non-metric variables.

\section{General teleparallel geometric theory of defects} \label{sec:gtpg-theory-of-defects}

Within the generalized teleparallel geometric framework that we have constructed, it is very natural to reproduce the defect hierarchy presented in the series of papers \cite{dewit1970i}–\cite{dewit1973iv}. In our earlier work \cite{adak-kok-2025}, where we were not yet able to explicitly identify the point defect, the Burgers vector, the generalized Burgers vector, and the Frank vector in a transparent manner, we attempted to bring the correct defect hierarchy to the forefront by employing generalized teleparallel geometry. Essentially, in that work we associated the non-Riemannian contribution in the decomposition of the full connection given in (\ref{eq:connec-decom}), namely the defect 1-form, with dislocations and disclinations as follows,
\begin{align} \label{eq:kusur-formu1}
L_{ab} = \underbrace{ K_{ab} + ( \imath_b Q_{ac} - \imath_a Q_{bc} ) e^c }_{dislocation} + \underbrace{ Q_{ab} }_{disclination} .
\end{align}

In the present work, while remaining faithful to the core idea developed in \cite{adak-kok-2025}, we introduce a reduced set of effective defect
variables constructed from selected vectorial and scalar sectors of torsion and non-metricity. As discussed in Sec.~2, it is essential to distinguish local defect-density fields from their corresponding integrated defect measures.

For the dislocation sector, our convention $*T^a=\alpha^a{}_{b}e^b$ identifies the full vector-valued torsion 2-form with the local dislocation-density tensor. The Burgers vector associated with an oriented surface $\mathcal{S}$ is the corresponding flux,
\begin{equation}
    b^a(\mathcal{S}) = \int_{\mathcal{S}}\alpha^a{}_{b}*e^b,
\label{burgers_flux}
\end{equation}
while the finite-circuit Burgers vector is the affine-development closure failure $B(p;C)$ defined in (\ref{burgers_affine}). Consequently, the torsion trace $T$ retained in the present reduced model is not identified with the complete Burgers measure; rather, as shown in (\ref{torsion_alpha_decomposition})-(\ref{T_S_dislocation}), it parametrizes the axial vectorial projection of the local dislocation-density tensor.

Similarly, let $\theta^{ba}$ denote the conventional local disclination-density tensor. The Frank vector associated with an oriented surface $\mathcal{S}$ is the integrated flux
\begin{equation}
    \Omega^a(\mathcal{S})
    = \int_{\mathcal{S}}\theta^{ba}*e_b .
\label{frank_flux}
\end{equation}
In \cite{adak-kok-2025} we introduced a tensorial correspondence between the disclination density and non-metricity. In the present reduced formulation, the three-component non-metricity covector $P$ is not
identified with the complete integrated Frank measure. Instead, it is retained as an effective Frank-type covector parametrizing the selected
vectorial disclination sector.
\begin{subequations} \label{eq:matchings-defects-gtpg}
    \begin{align}
\text{vectorial dislocation sector}
&\sim \text{effective Burgers-type covector},
& \tilde{b} &:= T ,
\\
\text{vectorial disclination sector}
&\sim \text{effective Frank-type covector},
& \tilde{\Omega} &:= P ,
\\
\text{point-defect sector}
&\sim \text{point-defect covector},
& \tilde{m} &:= Q ,
\\
\text{scalar torsional sector}
&\sim \text{additional scalar},
& S &:= \rho *1 ,
\\
\text{shifted translational defect sector}
&\sim \text{generalized Burgers-type covector},
& \tilde{B} &:= T+c_1P+c_2Q . \label{59e}
    \end{align}
   \end{subequations}
Here and in the remainder of the paper, the symbol "$\sim$" denotes an association between a defect sector and the effective geometric variable used to parametrize that sector; it does not denote equality with the conventional integrated Burgers or Frank measure. In particular, $\tilde{b}$ and $\tilde{\Omega}$ distinguish the effective local covectors of the present reduced model from the conventional Burgers and Frank vectors $b^a(S)$ and $\Omega^a(S)$ defined above.

The identifications in (\ref{eq:matchings-defects-gtpg}) define a deliberately reduced defect sector and are not intended to parametrize the most general local
dislocation and disclination density tensors. As shown in Sec.~2, a general dislocation density may contain a non-vanishing five-component symmetric-traceless sector represented by $\overset{(1)}{T^a}$; configurations requiring this sector therefore lie
outside the present truncation. Likewise, in the non-metricity sector we retain only the three-component $P$ and $Q$ sectors and set the seven-component $\overset{(1)}{Q_{ab}}$ and five-component
$\overset{(2)}{Q_{ab}}$ sectors to zero. Thus, configurations requiring these omitted tensorial sectors are outside the scope of the present reduced
model. In particular, Eqs.~\eqref{straight_dislocation_density}--%
\eqref{sym_traceless_dislocation} show that conventional straight screw and edge dislocations generally possess components in the omitted $\overset{(1)}{T^a}$ sector; the present variables therefore retain selected projections of their local densities rather than their full tensorial content.

The constants $c_1$ and $c_2$ in \ref{59e} are initially left free. If one wishes to recover the corresponding combination from the defect 1-form (\ref{eq:kusur-formu1}), substitution of the reduced forms (\ref{eq:burulma-ayris1}) and (\ref{eq:nonmetricity-ayris1}) gives $c_1=-1$ and $c_2=2/3$. Accordingly, in the following calculations we use
\begin{equation}\label{eq:BTPQ}
    \tilde{B}
    :=
    T-P+\frac{2}{3}Q .
\end{equation}
We emphasize that $\tilde{B}$ is a generalized Burgers-type covector of the present reduced model and is not identified with the translational part of a general
Euclidean holonomy.

Thus, throughout the remainder of the paper we use
$\tilde{b}:=T$, $\tilde{\Omega}:=P$, $\tilde{m}:=Q$, and $S:=\rho *1$ as effective local defect variables.
With these definitions, the reduced forms (\ref{eq:burulma-ayris1}) and (\ref{eq:nonmetricity-ayris1}) become
\begin{subequations} \label{eq:gtpg-defect-tors-nonmet}
\begin{align}
T^a &= \frac{1}{2} e^a \wedge \tilde{b} + \frac{\rho}{3} *e^a ,\label{eq:tors-defect1}\\
Q_{ab} &= \frac{3}{10} \left(\Omega_a e_b + \Omega_b e_a - \frac{2}{3} \delta_{ab} \tilde{\Omega} \right) + \frac{1}{3} \delta_{ab} \tilde{m} . \label{eq:nonmet-defect1}
\end{align}
\end{subequations}
We shall substitute these two equations into the Bianchi identities given in (\ref{eq:bianchi-identities}) (by setting $R^a{}_b=0$) and derive kinematic relations among the defects.

Accordingly, with the help of (\ref{eq:tors-defect1}), the second Bianchi identity (\ref{eq:bianchi-identities}ii) in general teleparallel geometry takes the form
\begin{align}
&\frac{1}{2} T^a \wedge \tilde{b} - \frac{1}{2} e^a \wedge d\tilde{b} + \frac{1}{3} d\rho \wedge *e^a \nonumber \\
& \qquad \qquad + \frac{1}{3} \rho (2Q^{ab} \wedge *e_b - Q \wedge *e^a + T^b \wedge *e^a{}_b) =0 .
\end{align}
If we now reinsert the equations (\ref{eq:gtpg-defect-tors-nonmet}) into this expression, it yields
\begin{align}
d\tilde{b} \wedge e^a + \frac{1}{3} \rho \tilde{b} \wedge *e^a -\frac{4}{3}\rho \tilde{\Omega} \wedge *e^a - \frac{2}{3} d\rho \wedge *e^a + \frac{2}{9} \rho \tilde{m} \wedge *e^a =0 .
\end{align}
This exterior-form equation corresponds to a tensor equation $S^a=0$ where $S^a$ is a 0-form giving rise to three scalar equations. These can be written in vector notation as
\begin{align} \label{eq:kinetik-eqn1}
\vec{\nabla} \times \vec{b} = - \frac{1}{3} \rho \vec{b} + \frac{4}{3} \rho \vec{\Omega} + \frac{2}{3} \vec{\nabla}\rho - \frac{2}{9} \rho \vec{m} .
\end{align}
Here we have used the relations $d\rho = (\vec{\nabla} \rho)_a e^a$ and $d\tilde{b} = (\vec{\nabla} \times \vec{b})_a *e^a$. 

Similarly, let us substitute equation (\ref{eq:nonmet-defect1}) into the third Bianchi identity (\ref{eq:bianchi-identities}iii) in general teleparallel geometry,
\begin{align}\label{62t}
&\frac{3}{10} \left[ \left( D\Omega_a \right) \wedge e_b + \Omega_a (De_b) + (a \leftrightarrow b) \right] \nonumber \\
& \qquad \qquad + \frac{2}{5} Q_{ab} \wedge \tilde{\Omega} - \frac{1}{5} \delta_{ab} d\tilde{\Omega} - \frac{2}{3} Q_{ab} \wedge \tilde{m} + \frac{1}{3} \delta_{ab} d\tilde{m} = 0 ,
\end{align}
where we use $D\Omega_a = \frac{1}{2} \Omega_b (\iota_a T^b) - \frac{1}{2} \iota_a d\tilde{\Omega} + D_{(a}\Omega_{b)}e^b$ and reinsert the relations given in (\ref{eq:gtpg-defect-tors-nonmet}) into (\ref{62t}). In this way, we obtain the following tensorial equality
\begin{align} \label{eq:kinemtik-eqns2}
&
\frac{3}{10}
D_{(k}\Omega_{a)}\epsilon^{k}{}_{bc}
+
\frac{3}{10}
D_{(k}\Omega_{b)}\epsilon^{k}{}_{ac}
\nonumber\\
&
-\frac{3}{20}
(\vec{\nabla}\times\vec{\Omega})_b\delta_{ac}
-\frac{3}{20}
(\vec{\nabla}\times\vec{\Omega})_a\delta_{bc}
+\frac{1}{10}
(\vec{\nabla}\times\vec{\Omega})_c\delta_{ab}
+\frac{1}{3}
(\vec{\nabla}\times\vec{m})_c\delta_{ab}
\nonumber\\
&
-\frac{3}{40}
b^k\Omega_a\epsilon_{kbc}
-\frac{3}{40}
b^k\Omega_b\epsilon_{kac}
-\frac{3}{40}
b_a\Omega^k\epsilon_{kbc}
-\frac{3}{40}
b_b\Omega^k\epsilon_{kac}
\nonumber\\
&
+\frac{3}{20}\rho\Omega_b\delta_{ac}
+\frac{3}{20}\rho\Omega_a\delta_{bc}
-\frac{1}{10}\rho\Omega_c\delta_{ab}
\nonumber\\
&
+\frac{9}{50}
\Omega^k\Omega_a\epsilon_{kbc}
+\frac{9}{50}
\Omega^k\Omega_b\epsilon_{kac}
=0 .
\end{align}
Equation (\ref{eq:kinemtik-eqns2}) is a three-index tensor equation,
$S_{(ab)c}=0$. Taking its trace and the remaining independent
contractions yields the following set of kinematic relations:
\begin{subequations} \label{eq:rot-point-defects}
\begin{align}
\vec{\nabla}\times\vec{m}
&=0, \label{eq:rot-ponit-def1}
\\
\vec{\nabla}\times\vec{\Omega}
&=\rho\,\vec{\Omega},\label{eq:rot-disklin1}
\\
D_{\langle a}\Omega_{c\rangle}
&=
\frac{1}{4}
\left(
b_a\Omega_c+b_c\Omega_a
\right)
-\frac{3}{5}\Omega_a\Omega_c
\nonumber\\
&\quad
-\frac{1}{6}
(\vec b\cdot\vec\Omega)\delta_{ac}
+\frac{1}{5}
(\vec\Omega\cdot\vec\Omega)\delta_{ac}. \label{symmetric_trace-free1}
\end{align}
\end{subequations}
Here angular brackets denote the symmetric trace-free part of the
covariant gradient, $D_{\langle a}\Omega_{c\rangle}
:= D_{(a}\Omega_{c)} -\frac{1}{3}\delta_{ac}D^b\Omega_b$. As a result, within the general teleparallel geometric theory of defects developed here, the kinematic equations hidden in the structure of the Bianchi identities are explicitly revealed in equations (\ref{eq:kinetik-eqn1}) and (\ref{eq:rot-point-defects}). Equation (\ref{eq:rot-disklin1}) shows that the effective Frank-type field belongs to the class of generalized Beltrami fields, whereas (\ref{symmetric_trace-free1}) determines the symmetric trace-free part of its covariant gradient in terms of the local defect variables $\vec b$ and $\vec\Omega$. Furthermore, the exterior-form counterpart of equation (\ref{eq:rot-ponit-def1}) is simply $d\tilde{m}=0$. Accordingly, for an arbitrary 0-form field $\phi(\vec{r})$, one may write $\tilde{m} = d\phi$, and the particular choice $\phi(\vec{r}) = \ln{\lambda(\vec{r})}$ corresponds to equation (33) of \cite{miri2002}. In this case, the main result of \cite{miri2002} can be rewritten here as $d*d\phi = N$, where $N=\rho_{exmt}*1$ is the extra-matter 3-form and $\rho_{exmt}$ is the extra-matter density 0-form. The total amount of extra-matter contained in a volume $V$ is then given by $\int_V \rho_{exmt}*1 = \int_V d*d\phi = \oint_{\partial V} *d\phi$, where Stokes’ theorem has been used in the last step. The final surface integral $\oint_{\partial V} *d\phi$ represents the net extra-matter flux through the closed surface enclosing the volume.

In addition, the scalar field $\rho=\mathrm{tr}\,\alpha$, representing the trace sector of the local dislocation-density
tensor, plays an important role, particularly in equations (\ref{eq:kinetik-eqn1}) and (\ref{eq:rot-disklin1}). Meanwhile, in a recent paper \cite{hirth2025}, the authors modified the traditional linear-elastic field of an edge dislocation based on symmetry and energy requirements. They showed that these requirements are satisfied by the addition of a new line force $\vec{f}$ perpendicular to $\vec{b}$. Accordingly, our new vector $\vec{\nabla}\rho$ may correspond to their new line-force vector. In the present paper, we do not want to over-emphasize this particular possibility.

In our general teleparallel geometric theory of defects, the Lagrange 3-form representing the free energy density can be written in its most general form as
\begin{align} \label{eq:lagrange-gtpg-defect1}
L &= \kappa_1 T \wedge *T + \kappa_2 S \wedge *S + \kappa_3 P \wedge *P + \kappa_4 Q \wedge *Q \nonumber \\
& \qquad \qquad \qquad + \kappa_5 P \wedge *Q + \kappa_6 T \wedge *P + \kappa_7 T \wedge *Q ,
\end{align}
where $\kappa_i$, $i=1,2,\cdots,7$, are coupling constants. For readers who are not familiar with the exterior-algebra formulation, it is useful to rewrite this expression in vector notation,
\begin{align} \label{eq:lagrange-gtpg-defect-vec}
L &= \big( \kappa_1 \vec{b} \vdot \vec{b} + \kappa_2 \rho^2 + \kappa_3 \vec{\Omega} \vdot \vec{\Omega} + \kappa_4 \vec{m} \vdot \vec{m} \nonumber \\
& \qquad \qquad \qquad + \kappa_5 \vec{\Omega} \vdot \vec{m} + \kappa_6 \vec{b} \vdot \vec{\Omega} + \kappa_7 \vec{b} \vdot \vec{m} \big) *1 .
\end{align}
One may also add terms of the form $(\vec{b} \times \vec{\Omega}) \vdot \vec{m} \, *1 \sim T \wedge P \wedge Q$ to the Lagrangian. However, such terms do not preserve parity, since they contain either a vector cross product in the vector representation or, equivalently, an even number (here zero) of Hodge maps in the exterior-algebra formulation. Higher polynomial order in the defect fields does not by itself
increase the differential order of the Euler--Lagrange equations. For example, cubic or quartic invariants constructed algebraically from $T$, $S$, $P$, and $Q$ still involve the same derivative order of the underlying geometric variables. In the present work we
restrict ourselves to the parity-even quadratic sector because it provides the minimal constitutive ansatz and permits a direct comparison with the quadratic general teleparallel Lagrangian considered in \cite{adak2023ijgmmp}. Higher-order interaction terms may be included in a more general effective theory, but they are not considered here.   

Furthermore, in order to compare the Lagrangian (\ref{eq:lagrange-gtpg-defect1}) with the Lagrange 3-forms used in the context of alternative theories of gravity in \cite{adak2023ijgmmp}, we derive the following relations
\begin{subequations}\label{q_terms_2}
\begin{align}
T \wedge *T &= T^a \wedge *T_a - (T^a \wedge e_b) \wedge *(T^b \wedge e_a) ,\\
S \wedge *S &= (T^a \wedge e_a) \wedge *(T^b \wedge e_b) ,\\
P \wedge *P &= Q^{ab} \wedge *Q_{ab} - (Q^{ab} \wedge e^c) \wedge *(Q_{ac} \wedge e_b) \nonumber \\
& \qquad \qquad - \frac{5}{9} Q \wedge *Q + \frac{2}{3} (Q \wedge e^b) \wedge *(Q_{ab} \wedge e^a) ,\\
P \wedge *Q &= \frac{2}{3} Q \wedge *Q - (Q \wedge e^b) \wedge *(Q_{ab} \wedge e^a) , \\
T \wedge *P &= - (Q^{ab} \wedge e_{ac}) \wedge *(T^c \wedge e_b) - \frac{2}{3} (Q\wedge e_a) \wedge *T^a \nonumber \\
& \qquad \qquad \qquad + (Q_{ab} \wedge e^b) \wedge *T^a ,\\
T \wedge *Q &= - (Q\wedge e_a) \wedge *T^a .
\end{align}
\end{subequations}
We can now state the following result. The Lagrangian given by equation (73) in \cite{adak2023ijgmmp} coincides with the Lagrangian (\ref{eq:lagrange-gtpg-defect1}) presented here, provided that the coupling constants $k_1,k_2,k_3$, $c_1,c_2,c_3,c_4,c_5$, $\ell_1,\ell_2,\ell_3$ used there are related to the present coupling constants $\kappa_i$ as
\begin{align}\label{k_c_l_constants}
k_1=\kappa_1, \quad k_2 = \kappa_2, \quad k_3 = -\kappa_1 , \quad c_1 = \kappa_3 , \quad c_2 = 0, \nonumber \\
c_3=-\kappa_3 , \quad c_4 = -\frac{5}{9} \kappa_3 + \kappa_4 + \frac{2}{3} \kappa_5 , \quad c_5 = \frac{2}{3}\kappa_3 - \kappa_5 , \nonumber \\
\ell_1 = - \kappa_6 , \quad \ell_2 = -\frac{2}{3} \kappa_6 -\kappa_7 , \quad \ell_3 = \kappa_6 .
\end{align}

We now consider the variational field equations of the model. Since the relations in (\ref{q_terms_2}) are algebraic identities, the Lagrangian (\ref{eq:lagrange-gtpg-defect1}) can be varied before imposing the reduced forms (\ref{eq:gtpg-defect-tors-nonmet}). We follow the first-order variational procedure of \cite{adak2023ijgmmp}. We work in an orthonormal frame, where the metric components $\delta_{ab}$ are fixed, $\delta\delta_{ab}=0$. Therefore, the variation of the metric is contained in the variation of the orthonormal coframe through $g=\delta_{ab}e^a\otimes e^b$. Accordingly, $e^a$ and $\omega^a{}_b$ are taken as independent geometric variables. The zero-curvature condition is imposed by a Lagrange multiplier
1-form $\chi^b{}_a$. Thus, in three dimensions we consider the constrained Lagrangian
\begin{equation}
L_{\mathrm{c}}
=
L
+
R^a{}_b\wedge\chi^b{}_a ,
\end{equation}
where $L$ is given in (\ref{eq:lagrange-gtpg-defect1}). The variations of the field strengths are $\delta T^a = D(\delta e^a) +
\delta\omega^a{}_b\wedge e^b$, $\delta Q_{ab} =
\delta\omega_{(ab)}$ and $\delta R^a{}_b = D(\delta\omega^a{}_b)$. Up to an exact form, the variation of the constrained Lagrangian can be written as
\begin{equation}
\delta L_{\mathrm{c}}
=
\delta e^a\wedge\tau_a
+
\delta\omega^a{}_b\wedge
\left(
\Sigma^b{}_a+D\chi^b{}_a
\right)
+
\delta\chi^b{}_a\wedge R^a{}_b ,
\end{equation}
where $\tau_a:=\frac{\delta L}{\delta e^a}$ and
$\Sigma^b{}_a:=\frac{\delta L}{\delta\omega^a{}_b}$.
Using the decomposition (\ref{q_terms_2}), these response forms can be written in the same basis as in \cite{adak2023ijgmmp},
\begin{align}
\tau_a
={}&
\sum_{i=1}^{3}k_i\,{}^{(i)}\tau_a[T]
+
\sum_{i=1}^{5}c_i\,{}^{(i)}\tau_a[Q]
+
\sum_{i=1}^{3}\ell_i\,{}^{(i)}\tau_a[QT],
\\
\Sigma^b{}_a
={}&
\sum_{i=1}^{3}k_i\,{}^{(i)}\Sigma^b{}_a[T]
+
\sum_{i=1}^{5}c_i\,{}^{(i)}\Sigma^b{}_a[Q]
+
\sum_{i=1}^{3}\ell_i\,{}^{(i)}\Sigma^b{}_a[QT],
\end{align}
where the coupling constants $k_i$, $c_i$, and $\ell_i$ are given explicitly in terms of $\kappa_i$ by (\ref{k_c_l_constants}). The elementary
coframe and connection response forms are those derived in \cite{adak2023ijgmmp}. Since the variations $\delta e^a$, $\delta\omega^a{}_b$, and $\delta\chi^b{}_a$ are independent, the field equations are
\begin{subequations}\label{field_eqs_a}
\begin{align}
\tau_a &=0,
\\
\Sigma^b{}_a+D\chi^b{}_a &=0,
\\
R^a{}_b &=0.
\end{align}
\end{subequations}
These are the coframe equation, the connection equation, and the zero-curvature constraint, respectively. The reduced conditions used in the defect classification are not imposed before the variation. After deriving the field equations, we restrict them to the sector $\overset{(1)}{T^a}=0$, $\overset{(1)}{Q_{ab}}=0$, $\overset{(2)}{Q_{ab}}=0$, or equivalently to the reduced forms (\ref{eq:gtpg-defect-tors-nonmet}). Thus, the present procedure corresponds to varying the general field equations first and restricting them to the reduced defect sector afterwards.

Let us emphasize once again that, as explicitly demonstrated in Ref.~\cite{adak2023ijgmmp}, a metric formulation of general teleparallel geometry is possible. This constitutes one of the strong aspects of the general teleparallel geometric theory of defects.

The final point to be discussed within the present framework concerns the relation of the torsional terms in (\ref{eq:lagrange-gtpg-defect1}) to the
classical strain energies of straight dislocations. For a pure dislocation sector, $P=Q=0$, and the torsional contribution to the free energy is $L_{\rm disl} = \kappa_1 T\wedge *T + \kappa_2 S\wedge *S$. For a locally straight dislocation, (\ref{T_S_dislocation}) gives $T^\sharp
=
\varrho_\ell(\vec{\xi}\times\vec{b}_{\rm B})$ and $S
=
\varrho_\ell
(\vec{b}_{\rm B}\cdot\vec{\xi})*1$,
where $\vec{b}_{\rm B}$ denotes the conventional Burgers vector. Let $\varrho_\ell$ be a regularized areal line-density profile normalized by $\int_A \varrho_\ell\,dA=1$, and define $I_\ell :=
\int_A \varrho_\ell^2\,dA$.

For a straight edge dislocation,
$\vec{b}_{\rm B}\perp\vec{\xi}$, and therefore $S=0$. The contribution of the reduced torsional sector to the energy per unit length is
\begin{equation}
\frac{E_{\rm edge}}{L_z}
=
\kappa_1 b_{\rm B}^2 I_\ell .
\end{equation}
For a straight screw dislocation,
$\vec{b}_{\rm B}\parallel\vec{\xi}$, and therefore $T=0$. In this
case,
\begin{equation}
\frac{E_{\rm screw}}{L_z}
=
\kappa_2 b_{\rm B}^2 I_\ell .
\end{equation}
The corresponding classical line energies in isotropic linear elasticity are  $\frac{E_{\rm screw}}{L_z}
= \frac{G b_{\rm B}^2}{4\pi}
\ln\frac{R}{r_0}$ and $\frac{E_{\rm edge}}{L_z}
= \frac{G b_{\rm B}^2}{4\pi(1-\nu)}
\ln\frac{R}{r_0}$ \cite{Hirth_Lothe_1982}.
Hence, for the same regularized line-density profile, the couplings of the retained torsional sectors may be calibrated as
\begin{equation}
\kappa_2 = \frac{G}{4\pi I_\ell}
\ln\frac{R}{r_0},
\qquad
\kappa_1 = \frac{G}{4\pi(1-\nu)I_\ell}
\ln\frac{R}{r_0}.
\end{equation}
These calibrated values should be regarded as effective parameters for the chosen regularized defect profile and specimen geometry, rather than as universal material constants. Indeed, $[\varrho_\ell]=L^{-2}$ and $[I_\ell]=L^{-2}$. Since $[G]=E\,L^{-3}$, the above relations
give $[\kappa_1]=[\kappa_2]=E\,L^{-1}$, so that integration of $L_{\rm disl}$ over a volume has the dimension of energy. These relations should be understood as a calibration of the torsional sectors retained in the present reduced model. As discussed in Sec.~2, a general straight screw or edge dislocation
also contains the omitted symmetric-traceless torsion sector $\overset{(1)}{T^a}$. Therefore, the expressions above do not represent a reconstruction of the complete classical dislocation energy, but only its correspondence with the retained effective
projections.

\subsection{Explicit defect configurations}

Following the geometric construction used for cylindrically symmetric screw-dislocation distributions \cite{Yavari_Gorielli_2012}, \cite{Yavari_Goriely_2014}, we consider a regularized straight screw dislocation along the $z$-axis. We introduce a regularized line-density profile
$\varrho_\ell(r)$ satisfying $2\pi\int_0^\infty \varrho_\ell(r)\,r\,dr=1$ and define $f(r):=\int_0^r\varrho_\ell(s)s\,ds$.
Thus, $f(\infty)=1/(2\pi)$.

We choose the coframe
\begin{equation}
e^1=dx,\qquad
e^2=dy,\qquad
e^3=dz+b_{\rm B}f(r)d\phi ,
\end{equation}
and the Weitzenb\"ock gauge $\omega^a{}_b=0$.
Consequently, $Q_{ab}=0$ and $R^a{}_b=0$ while the only non-vanishing torsion component is
\begin{equation}
T^3
=
b_{\rm B}\varrho_\ell(r)e^{12}.
\end{equation}
Integrating over the complete transverse section, the Burgers measure is therefore
\begin{equation}
    \int_A T^3
=
b_{\rm B}\int_A\varrho_\ell\,dA
=
b_{\rm B}.
\end{equation}
Equivalently, in the limit of a large circuit $C$ enclosing the
complete defect profile,
\begin{equation}
    \oint_C e^3=b_{\rm B}.
\end{equation}
Thus, the conventional Burgers vector is reproduced by the integrated torsion flux.

For this configuration, the irreducible torsional variables are $T=\iota_aT^a=0$ and $S=e_a\wedge T^a = b_{\rm B}\varrho_\ell(r)*1$. 
Hence,
\begin{equation}
    \rho=b_{\rm B}\varrho_\ell(r).
\end{equation}
This agrees with the general relations in (\ref{T_S_dislocation}) for 
$\vec b_{\rm B}\parallel\vec\xi$.

However, the full torsion of this screw dislocation is not equal to its scalar irreducible part alone. In particular, $\overset{(1)}{T^a}\neq0$. Therefore, the standard screw dislocation is not completely contained in the reduced torsional sector (\ref{eq:burulma-ayris1}). The reduced model retains its scalar projection $S$, while the symmetric-traceless part is omitted.

We next give a non-trivial configuration that lies entirely within the reduced sector. Consider the punctured space
\begin{equation}
    M=\mathbb{R}^3\setminus\{x=y=0\},
\end{equation}
with $e^1=dx$, $e^2=dy$, $e^3=dz$
and define the globally defined closed 1-form $\eta :=
\frac{x\,dy-y\,dx}{x^2+y^2}$, which locally can be written as $\eta=d\phi$. Although $d\eta=0$ on $M$, $\eta$ is not globally exact and satisfies $\oint_C\eta=2\pi $ for a closed curve $C$ surrounding the $z$-axis.

Let the affine connection be
\begin{equation}
\omega^a{}_b
=
\frac{q}{3}\delta^a_b\,\eta ,
\end{equation}
where $q$ is a constant. Since $d\eta=0$ and
$\eta\wedge\eta=0$, its curvature vanishes identically,
$R^a{}_b=0$. The corresponding torsion and non-metricity are
\begin{equation}
T^a
=
\frac{q}{3}\eta\wedge e^a,
\qquad
Q_{ab}
=
\frac{q}{3}\delta_{ab}\eta .
\end{equation}
The effective variables are therefore
\begin{equation}\label{eff_variables}
T=-\frac{2q}{3}\eta,
\qquad
S=0,
\qquad
P=0,
\qquad
Q=q\eta .
\end{equation}
These fields satisfy the reduced forms (\ref{eq:burulma-ayris1}) and (\ref{eq:nonmetricity-ayris1}) exactly.
Thus, $\overset{(1)}{T^a}=0$, $\overset{(1)}{Q_{ab}}=0$ and $\overset{(2)}{Q_{ab}}=0$.

The shifted translational defect covector defined in (\ref{eq:BTPQ}) becomes $\tilde B
= T-P+\frac{2}{3}Q = 0$. Nevertheless, both torsion and non-metricity are non-vanishing. Thus, this configuration provides a simple combined dislocation--point-defect field in which the two retained vectorial sectors compensate in $\tilde B$. Moreover, $d\tilde b=0$, $d\tilde m=0$, $\tilde\Omega=0$ and $\rho=0$. Therefore, the kinematic relations (\ref{eq:kinetik-eqn1}) and (\ref{eq:rot-point-defects}) are satisfied
identically.

This configuration can also satisfy the variational field equations. For the coupling choice $\kappa_7=\frac{4}{3}\kappa_1$, $\kappa_4=\frac{4}{9}\kappa_1$, $\kappa_5=\frac{2}{3}\kappa_6$ the Lagrangian can be written as
\begin{equation}
    L
=
\kappa_1 U\wedge *U
+\kappa_2 S\wedge *S
+\kappa_3 P\wedge *P
+\kappa_6 P\wedge *U,
\qquad
U:=T+\frac{2}{3}Q.
\end{equation}
For the configuration (\ref{eff_variables}), one has
$U=0$, $P=0$ and $S=0$. Hence both the Lagrangian and its first variation vanish, and therefore $\tau_a=0$ and $\Sigma^b{}_a=0$. Taking also $\chi^b{}_a=0$, the three variational equations (\ref{field_eqs_a}) are satisfied:
\begin{equation}
    \tau_a=0,
\qquad
\Sigma^b{}_a+D\chi^b{}_a=0,
\qquad
R^a{}_b=0.
\end{equation}
Hence, this configuration provides an explicit exact solution of the reduced vacuum model for the above subfamily of coupling constants.

\section{Results and discussion}

In this work we have reformulated the geometric theory of crystalline defects within the framework of general teleparallel geometry. All constructions were performed in the language of exterior calculus using coordinate-independent orthonormal frames, so that the structural content of the theory is separated from coordinate-dependent artifacts.

We first clarified the geometric description of finite elastic deformations by distinguishing the material manifold from the spatial configuration. The deformation is described by a regular map
$\varphi:\mathcal{B}\rightarrow\mathcal{S}$, while the constitutive response is formulated within nonlinear elasticity. In the presence of distributed defects, the deformation map may remain regular, while
the defect content is encoded in the intrinsic geometry of the material manifold.

In contrast to the mainstream metric-affine identification of defects---where torsion represents dislocations, curvature represents disclinations, and non-metricity represents point defects---we adopted a curvature-free setting, $R^a{}_b=0$, and reorganized the defect variables as irreducible components of torsion and non-metricity. Within this curvature-free framework, the torsion trace is retained as an effective vectorial Burgers-type sector, the vectorial
non-metricity covector $P$ as an effective Frank-type sector, and the non-metricity trace $Q$ as a point-defect sector. The scalar torsional component represents the trace projection of the local
dislocation-density tensor and is retained as an additional scalar degree of freedom of the reduced model. These variables represent selected projections of the complete defect densities, and configurations requiring the omitted irreducible sectors lie outside the present reduced model. This reorganization avoids the particular algebraic overlap that arises when
torsion-dependent contributions contained in the full curvature are simultaneously interpreted as an independent defect sector. It does not, however, remove the intrinsic origin dependence of the translational part of a general Euclidean holonomy in the presence of a nontrivial rotational part. The present construction should therefore be understood as a reduced local classification of effective defect variables rather than as a resolution of this general holonomy property. Compared with our previous formulation \cite{adak-kok-2025}, the present construction also distinguishes explicitly between local defect-density fields and integrated Burgers and Frank measures, and identifies the retained $T$, $S$, $P$, and $Q$ sectors separately.

The Bianchi identities were shown to encode the kinematic relations among the defect variables, making explicit the differential constraints governing their mutual consistency. In particular, the effective Frank-type field satisfies a generalized Beltrami relation, while the symmetric trace-free part of its
covariant gradient is constrained by the remaining defect variables. The point-defect covector is locally closed. General teleparallel geometry also admits a metric formulation \cite{adak2023ijgmmp}. Quadratic Lagrangians constructed solely from torsion and non-metricity can lead to second-order field equations in the teleparallel metric formulation considered here, without invoking curvature-squared terms \cite{adak2023ijgmmp}. We also derived the constrained first-order variational equations by taking the coframe and affine connection as independent geometric
variables and imposing $R^a{}_b=0$ through a Lagrange multiplier. The reduced defect conditions were imposed only after deriving the general field equations.

The torsional part of the quadratic free energy was compared with the classical line energies of straight screw and edge dislocations. This comparison gives effective values of $\kappa_1$ and $\kappa_2$
that depend on the regularized defect profile and specimen geometry. Two explicit configurations were also considered. The regularized screw-dislocation example reproduces the conventional Burgers measure
through the integrated torsion flux and shows explicitly that a standard screw dislocation also contains the omitted $\overset{(1)}{T^a}$ sector. A second curvature-free configuration with non-vanishing torsion and non-metricity lies entirely within the
reduced sector and provides an exact vacuum solution for the specified subfamily of coupling constants.

Beyond its application to crystalline defects, the present construction illustrates how non-Riemannian geometry can be reorganized into a curvature-free constrained field system with a well-defined variational structure. From this perspective, defect variables may be interpreted as emergent geometric field strengths rather than auxiliary material descriptors. In this sense, general teleparallel geometry provides a conceptual bridge between continuum defect theory and non-Riemannian field-theoretic formulations.

Several directions remain open. Solving the variational field equations for more general defect distributions and boundary conditions would allow the corresponding material and spatial geometries to be determined explicitly and compared with experimentally observed defect fields. Furthermore, symmetric teleparallel geometry—where non-metricity alone is present—may offer an alternative minimal setting, possibly endowed with a natural gauge structure \cite{Adak2026-history-stpg}. Exploring this possibility could further clarify the structural role of gauge principles in geometric defect theory.
  \\

\noindent 
{\bf Acknowledgements:} M. A. is supported by the TUBITAK Grant No 124F325. The authors thank the anonymous referees for their careful reading of the manuscript and helpful comments.\\

 
 \noindent
{\bf Data Availability Statement:} No data associated in the manuscript.

\end{document}